\documentstyle[12pt,prb,aps]{revtex}
\begin{document}
\draft
\title{Localization of Quantum States at the Cyclotron Resonance.}
\author{V. Ya. Demikhovskii, D. I. Kamenev}
\address{Nizhny Novgorod State University, Gagarin ave. 23, 
Nizhny Novogorod, 603600, Russia}
\address{E-mail: demi@phys.unn.runnet.ru}
\maketitle
\bigskip
\begin{abstract}
Tunneling of a quantum particle to the classically inaccessible
region in the intrinsically degenerated system is investigated
here by means of quasienergy eigenstates. The exact resonance
$\delta\omega=\omega-\omega_c=0$ and near resonance
$\delta\omega\ne 0$ cases
are explored numerically. It is shown that in both cases
all quantum states are localized. Correspondence
of the quantum dynamical barriers to the classical invariant
curves is demonstrated. The phenomena considered in this article
can be observed when an ultrasound wave propagates perpendicularly
to a magnetic field and interacts with a 2D electron gas
in a semiconductor heterostructure.

\end{abstract}
\bigskip

In this paper we treat the role of quantum dynamical barriers in
localization of quantum states at the cyclotron resonance.
Dynamical barriers in quantum systems are counterparts of
classical invariant tori (or KAM  tori) which block diffusion of
a particle along the phase space.
Similarly, in quantum case dynamical barriers inhibit the diffusion
of a wave packet in a Hilbert space
that leads to localization of quantum states.
This phenomenon was previously analysed for the kicked
rotator model [1, 2] and the nonlinear oscillator in a monochromatic
radiation field [3].
However quantum dynamics was explored maily for the systems,
whose classical counterparts are accidental degenerate and the
conditions of the KAM  theorem are satisfied.
The dynamics of the intrinsic degenerate system
(kicked simple harmonic oscillator) was explored
by Berman, Rubaev, Zaslavsky [5] under the
condition of strong and weak chaos of the classical limit.
In this work by
using quasienergy eigenstates we treat tunneling through
the dynamical barriers and localization of quantum states
in an intrinsically degenerated system where the KAM  theorem
is invalid.

Our system is a charged particle subject to both a uniform
magnetic field and a field of a longitudinal monochromatic wave,
propagating perpendicularly to the magnetic field direction.
The phase space of the classical system at the condition
of the cyclotron resonance
$\omega=\omega_c$, where $\omega$ and $\omega_c$ are the wave
and cyclotron frequencies, consists of the infinit number of cells
comprising closed classical orbits and separated by a
separatrix lattice [4].
In the classical case in the resonance approximation a particle has 
no
possibility to penetrate through a separatrix from one cell to 
another.
In this paper we will show that in the quantum case
the dynamical barriers become transparent for the wave packet even
in the resonance approximation. 

The theory of nonlinear resonance in the intrinsically degenerated
system was developed by G.A.Luna-Acosta and the authors
in Ref. [6]. It was
shown that the Hilbert space of the discussed system breaks up into
quantum resonance cells.
The boundaries of these cells in the quasiclassical limit correspond 
to
the the separatrices in the classical phase space. In this paper we
numerically investigate the role of these "quantum separatrices" in
localization and tunneling phenomena at the cyclotron resonance.

The Hamiltonian of the considered system has the form
\begin{equation}
\hat H=\frac{(\hat{\bf p}-{e\over c}{\bf A})^2}{2m}+V_0\cos(kx-\omega 
t)=
\hat H_0+V(x,t),
\end{equation}
where $e$ and $m$ are the charge and mass of a particle,
$c$ is the light velocity,
$\bf\hat p$ is the momentum,
$k$ and $V_0$ are the wave vector and amplitude of the perturbation.
It is convenient to choose the gauge of {\bf A}  in the form
{\bf A} $=$ (0,H$x$,0) in order
to produce the magnetic field {\bf H} along the z-direction.
In this  gauge the momentum $p_y$ is an integral of motion, hence
we have to determine the dependence of the wave function only on
two variables, $x$ and $t$.

Since the perturbation is periodic in time, the Floquet theory can
be used for description of time evolution of the system in terms
of quasienergy (QE) spectra and  QE eigenfunctions $\psi_q(x,t)$.
The QE eigenfunctions are eigenstates of the evolution operator for
one period of oscillation of the external field $T=2\pi/\omega$.
They can be defined by (see for example Ref. [7])
 \begin{equation}
 \psi _q(x,t)=\exp(-\frac{iE_qt}\hbar)\sum_n\sum_s A_{n,s}^q
 \psi_n(x)\exp(-is\omega t),
 \end{equation}
 where $E_q$ is the QE eigenvalues, $\psi_n (x)$ is the nth
 eigenfunction of a simple harmonic oscillator,  $\hbar$ is the 
Plank's
 constant. After substituting Eq. (2) into the time-dependent
  Schr\"odinger equation, we obtain the system of the uniform
  algebraic equations:
  \begin{equation}
 (E_q-\hbar\omega_c n+\hbar\omega s)A_{n,s}^q=
 \sum_m V_{n,m}(A_{m,s+1}^q+A_{m,s-1}^q),
 \end{equation}
 where the quasienergy $E_q$ is counted from the ground state
 $\hbar\omega_c/2$, the matrix elements $V_{n,m}$ was defined in Ref. 
[6].

First let us consider the exact resonance case where
$\delta\omega=\omega_c-\omega = 0$.
If $V_0=0$ each definite value $n-s$ in Eq. (3) corresponds to an
infinitely degenerated QE level and these levels are separated by the
distance $\hbar\omega$. When $V_0\ne 0$ the degeneracy is broken and 
each
initially degenerated level splits into QE spectrum.
All the spectra are identical, and the
distance between the extreme levels in one spectrum has the order of
$V_0$. If $V_0\ll\hbar\omega$ we can, at first order,
neglect the interaction between the individual spectra and treat
one of them.

 QE eigenvalues $E_q$ and eigenfunctions $A_{n,s}^q$ were calculated
 in Ref. [6] for resonant ($\omega=\omega_c$) and near resonant
 ($\delta\omega\ne 0$) case
 by the perturbation theory assuming  $V_0/\hbar\omega<<1$ and
 $\hbar\delta\omega n\ll\hbar\omega$ for all studied values of n.
 In the first order approximation Eq. (3) yields

 \begin{equation}
 (E_q-\hbar\delta\omega n)A_n^q=V_{n,n+1}A_{n+1}^q+
 V_{n,n-1}A_{n-1}^q,
 \end{equation}
 where $A_n^q\equiv A_{n,n}^q$. The matrix elements $V_{n,n\pm 1}$
 are symmetric, real
and for $n\gg 1$ can be expressed via the Bessel functions of the 
first
order
\begin{equation}
V_{n,n+1}={V_0\over 2}\sqrt{{n\over n+1}}e^{-\frac 
h4}J_1(\sqrt{2nh}),
\end{equation}
where $h=(ka)^2$, $a=({\hbar c\over eH})^{1/2}$ is the magnetic 
length.
When $\delta\omega=0$, the
amplitude of the wave $V_0$, which is measured in units of
$\hbar\omega$, does not influence the eigenvectors $A_n^q$
but changes only the energy scale, hence the
effective Plank's constant $h$ is the only parameter in this problem.
The oscillating matrix elements $V_{n,n+1}$ define the quantum
resonant cells --- relatively independent dynamical regions in the
Hilbert space [6]. Each quantum cell corresponds to two classical 
cells
(in action-angle variables).

When $\delta\omega\ne 0$ and $V_0=0$ according to Eq. (4) the levels 
are
separated from each other by the distance $\hbar\delta\omega$. If
$V_0\ne 0$ interaction between the levels appears, however,
due to Eq. (4) this interaction is essential only within the region
$n\hbar\delta\omega\sim V_0$,
whereas the levels satisfied the condition
$n\hbar\delta\omega\gg V_0$ practically are
not affected by the perturbation.

Let us discuss the main features of the QE spectrum which 
substantially
define dynamics of the system.
In the exact resonance case the spectrum $E_q$ is almost equidistant 
near
the top and the bottom at the distance $\hbar\tilde\omega$ [6].
The frequency $\tilde\omega$ in the quasiclassical limit is identical
to the frequency of small oscillations near the center of the 
resonance
in the phase space. In order to derive $\tilde\omega$ for the near
resonance case, let us expand $f(n)\equiv V_{n,n+1}$ near some
point $n_0$, defined below, into Taylor series and introduce
the differential operator by
\begin{equation}
A_{n+1}^q+A_{n-1}^q={d^2\over dn^2}A_n^q+2A_{n}^q.
\end{equation}
Then the finite-difference Shr\"odinger equation (4) takes the form
 \begin{equation}
 \left[ {\cal E}_q-(n-n_0)(\hbar\delta\omega+2f'(n_0))-2f(n_0) 
\right]
 A_n^q=f(n_0){d^2\over dn^2}A_n^q+f''(n_0)(n-n_0)^2A_n^q,
 \end{equation}
 where ${\cal E}_q=E_q-\hbar\delta\omega n_0$. If the condition
 \begin{equation}
\hbar\delta\omega=-2f'(n_0)
 \end{equation}
is satisfied equation (7) takes the simple harmonic oscillator form
with the center of oscillations $n_0$ defined by Eq. (8).
According to Eq. (8), when
$\delta\omega\ne 0$ there is only a finite number of the resonances.
If $V_0$ is small, i.e. $V_0<\hbar\delta\omega$, the unperturbed 
levels
$n\hbar\delta\omega$ are not coupled by the perturbation
(see Eqs. (4) or (8)) and the resonance disappears.
The same is true for the classical case.
In the limit $h \rightarrow 0$, $n\rightarrow \infty$, and finite
$nh=I$, the positions of the resonances and the frequency,
defined by Eqs. (8) and (7), correspond to the fixed points and
frequency of small oscillations near the center of the
resonance in the phase space [8].

 An arbitrary solution of the time-dependent Schr\"odinger equation
with the Hamiltonian (1) can be expanded in the simple harmonic
oscillator basis
\begin{equation}\psi (x,t)=\sum_n C_n(t)\psi_n(x)exp(-iE_nt/\hbar),
\end{equation}
were $E_n$ is the energy of nth Landau level.
Coefficients $C_n(t)$ determine the time development
of the probability distribution defined by $|C_n(t)|^2$.
Instead of integration of the time-dependent Schr\"odinger equation
we express the coefficients $C_n(t)$ in terms of the Green function
$G_{n,n'}$
\begin{equation}
C_n(t)=\sum_{n'}G_{n,n'}(t,t_0)C_{n'}(t_0),
\end{equation}
which can be determined by using the complete set of the
QE eigenstates
\begin{equation}
G_{n,n'}(t,t_0)=\sum_q A_n^qA_{n'}^qe^{-iE_q(t-t_0)/\hbar}.
\end{equation}

Thus, we have the following procedure for
computing the probability distribution at time $t$ which is used
below. First, by numerical solving the finite-difference
Schr\"odinger equation (4) we compute the QE eigenfunctions $A_n^q$ 
and
the QE eigenvalues $E_q$, then using the QE eigenstates we compute 
the
Green function (11), and finally we use the Green function for
calculating coefficients $C_n(t)$.
Since the dependence of the coefficients $C_n(t)$ on time $t$
in Eqs. (10), (11) is explicit, they can be easily computed
at arbitrary time $t$.

We have performed the numerical experiments in order to explore
how the dynamical barriers in the intrinsically degenerated system
(web tori [9]) influence the evolution of quantum states with various
initial conditions at the cyclotron resonance.
The system consists of the weakly interacting quantum resonant cells 
[6]
defined by the matrix elements $V_{n,n+1}$. Our purpose is to
treat electron tunneling between the cells.

As mentioned above, the QE states completely define the dynamics of
the system. If the initial state is any QE eigenstate,
i.e. $C_{n'}(t_0)=A_{n'}^{q'}$ , then in accordance with Eqs. (10) 
and (11)
the probability distribution does not depend on time $t$. In 
particular,
if the initial wave function $A_{n'}^{q'}$ is localized in one cell
then it will be localized in the initial cell for any time $t$.
The localized QE eigenfunctions correspond to the
eigenvalues near the top and bottom of the QE spectrum.
By employing  a coherent state representation (Husimi function) we 
can
show that the localization of a wave packet in the Hilbert space
entails the localization of the Husimi function in the phase space.
The contour plots of two Husimi functions corresponding to the
extream eigenvalues from the top and bottom of the QE spectrum are 
plotted
in Fig. 1 (a) (the upper part of the figure corresponds to the top 
eigenvalue
and the lower part corresponds to the bottom one).
The symmetry of these Husimi functions can be explained from Eq. (4)
which is invariant under the transformation
\begin{equation}
E_q \rightarrow -E_q, \qquad A_n^q \rightarrow (-1)^nA_n^q,
\end{equation}
corresponding to the transformation $x \rightarrow -x$
(or $\omega t \rightarrow \omega t+\pi$) in Eq. (2).
The contour plots of the Husimi functions correspond to the classical
orbits in two cells of the phase space which are shown in Fig. 1 (b).
Another kind of initial state is an excited Landau level
$C_n(0)=\delta_{n,n_0}$. In this situation, the initial wave packet
contains all the QE eigenstates and one can expect
that such a state will be extending most intensively, tunneling to 
other
resonant cells.

In our calculations the system involved
432 Landau levels which formed 7 resonance cells in the Hilbert 
space,
shown in the upper part of Fig. 2.
The initial state in the form $C_n(0)=\delta_{n,n_0}$
was placed in the center of the first cell. After the
time $\tilde T\sim 2\pi / \tilde\omega$ the wave
packet spreaded over the initial cell. In parallel with the fast
dynamics within the initial cell a slow process of
propagating the probability distribution to the subsequent cells took
place. The characteristic time of the probability distribution 
spreading
over all the considered cells
has the order of $2\pi / \omega_{min}$, where $\omega_{min}$
is the minimal distance between the QE eigenvalues in the QE spectrum
calculated for seven cells.
Two snapshots of evolution of the probability distribution
$|C_n|^2$ (in a logarithmic scale) as a function of the Landau number 
$n$
are shown in Figs. 2 (a), (b). It is seen that
the probability distribution in average
decreases exponentially, but sharp decay occures only
at the boudaries of the cells
that verifies our assumption of the role of these boudaries as
dynamical barriers to the probability flow.

The probability distribution successively penetrates through the 
dynamical
barriers from one cell to the subsequent ones, travelling along the
Hilbert space. After filling some cell the wave packet reflects
from the barrier and interferes
(in the sixth cell in Fig. 2 at the moment (a) $t=4*10^5$ and
(b) $t=7*10^5$, time is measured in units of $T=2\pi / \omega$).
As a result, oscillations of the probability distribution appear.
These oscillations can be explained also from another point of view.
Let us consider tunneling of some initial wave packet, concentrated
on a level with Landau number $n$ in an initial cell,
to another level with number $n'$, situated in another cell.
The main contribution to the Green function (11) is given by
the QE eigenfunctions with those numbers $q$, for
which both $A_n^q$ and $A_{n'}^q$ are sufficiently large or,
in other words,
any of these QE eigenfunctions must occupy two (or more) resonant 
cells.
Such eigenvectors correspond to the eigenvalues near the center
of the QE spectrum, i.e. the point $E_q=0$ [6].
As follows from our calculations, the most delocalized QE
eigenfunctions at the same time turn out to be
most oscillating (by module) functions of $n$.
The probability distribution in Fig. 2 also oscillates
with the minimal period. It is interesting that the
probability distribution in average is greater near the boundaries of 
the
cells (by 1.5 - 3 orders in Fig. 2). Since these boundaries
correspond to the separatrices [6] (web tori), this quantum 
phenomenon is
akin to the diffusion of a classical particle along the separatrice 
lattice
within exponentially small stochastic regions
in the phase space when nonresonant terms
in the classical Hamiltonian are taken into consideration [9].

The probability distribution in Fig. 2 is approximately the same
within a cell, hence it is reasonable to consider average
probabilities, defined by
$P_i(t)=\sum_{n_i}|C_{n_i}(t)|^2$, where $n_i$ takes the values
within the $i$th resonant cell. The evolution of $P_1(t)$ and 
$P_2(t)$
(in the first and second cells) is presented in Fig. 3
(the time axis is plotted in a logarithmic scale).
For the calculations we used the system with the same
parameters as in Fig. 2 involving seven resonance cells.
We suppose that the first two cells do not "feel" the limited
boundary conditions because the probability
distribution at the boundary of the system (the seventh cell)
is exponentially small and including of more number of cells does not
affect $P_1(t)$ and $P_2(t)$. The probability distribution in
Fig. 3 does not leak out to the subsequent cells in the limit
$t\rightarrow\infty$,
hence we can conclude that the quantum states are localized.
The localization of the quantum states
suggests a conjecture (Ref. [6]) about discreteness of the QE 
spectrum
for the unbounded system when $n\rightarrow\infty$.
Time-averaged probabilities $\overline{P_i(t)}$ versus cell number
for two values of the effective Plank's constant
$h$ are presented in Fig.~4.
In the logarithmic scale the exponential behaviour of these functions
is fairly evident. As one can see in Fig. 4 the relation
$\overline{P_2(t)}/\overline{P_1(t)} $ has the order of $10^{-2}$.
It should be stressed that the sharp decay of the probability 
distribution
occures not everywhere, but only at the boundaries of the quantum 
resonance
cells. Thus, we deal with the new kind of localization, namely,
localization over the quantum resonant cells. The localization length
in the resonance approximation
does not depend on the amplitude of the wave $V_0$. It is defined by
the only parameter $h=(ka)^2$ which determines the behaviour of
the Bessel function in the matrix elements in Eq. (5).
Our numerical experiments have shown that slope of
the curve on Fig. 4, defining the localization length, in general
increases with increasing $h$. However, the decay of the
probability distribution
at different boundaries between the cells changes randomly.
The nature of this randomness is discussed below.

Next effect, which we consider here, follows from discreteness of the
quantum number $n$ which labels Landau states in the
Hilbert space. If we choose the parameter $h$ so that
\begin{equation}
V_{n_0,n_0+1}\sim J_1(\sqrt{2n_0h})=0
\end{equation}
then in accordance with Eq. (4) transitions through the level $n_0$
are blocked and the Hilbert space is devided into two disjoint parts.
If we will change the parameter $h$, the condition (13)
will be successively satisfied for the $n_0+1$-th, $n_0+2$-th,
\dots levels. Thus the tunneling probability through the dynamical
barrier goes periodically to zero.
Let us define the penetration coefficient by 
$P=\sum_{n>n_0}|C_n(T)|^2$,
where $n_0$ is located at the boundary between
the second and the third cells, and place the initial state in the 
form
$C_n(0)=\delta_{n,n'}$ at the center of the second cell .
The oscillations of $P$ with changing the parameter $1/h\sim H$ are
shown in Fig. 5, where the above-mentioned zeroth values of the 
penetration
coefficient have been replaced by small numbers.
The oscillations have the period $\Delta H = 2\hbar ck^2 / eb^2$,
where b is a root (the second one in our case) of the Bessel function 
(5).
The amplitude of the oscillations decreases with decreasing $h$,
that does not contradict to the quasiclassical
formula $P\sim \exp(-\alpha/h)$. The definition of more explicit form
of the dependence of the tunneling probability $P$ on $h$ is
complicated by the wild oscillations of the dynamical barrier
permeabilities as a function of $h$.

The dynamical barriers become impermeable for definit values of $h$ 
only
in the resonance approximation when the condition
$\tilde\omega\ll\omega$ is satisfied and we take into account only
transitions between the nearest levels.

Next we present the results of culculations for the near resonance 
case
$\delta\omega\ne 0$ which are illustrated in
Fig. 6 where the probability distribution spreading after the time
$t=10^6$ is shown. There is a significant difference between
the near resnance and exact resonance cases.
Though the matrix elements (5) are the same as in
the exact resonance case, the additional term $n \hbar\delta\omega$ 
in (4)
deforms the resonance cells with small $n$'s and destroys the
cells with large $n$'s. That is indicated in Fig. 6 where the
boundary of the probability distribution spreading, pointed out by
the sharp decrease of this quantity, does not correspond to the
boundary of the second cell in the case
$\delta\omega=0$, marked by the second arrow on the
figure. As follows from calculations with decreasing $V_0$
(or increasing $\delta\omega$) the boundary of
the probability distribution spreading decreases.
The same is characteristic for the classical acidentally degenerate
system [8] where in the case $\delta\omega\ne 0$ the resonance 
occures
only in several resonance cells in the phase space and
the resonances are separated from each other by invariant curves.
According to equation (8), there is only
one resonance at choosen parameters, and the probability distribution
does not spread via the cells, no matter
how long we observe the dynamics. This is natural because,
as was shown above, the wave with a small amplitude $V_0$ does
not affect the states with large Landau numbers $n$. That
can be also understood from the expression for the Green function 
(11).
As one can see from Eq. (4), all the QE eigenstates with Landau 
numbers
$n$ from the region $n \hbar\delta\omega\gg V_0$ are localized,
$A_n^q=\delta_{n,q}$, with the characteristic
localization length $\Delta n\sim V_0/\hbar\delta\omega$ (that
can be shown for example in the quasiclassical limit).
On the other hand in order for any state to
evolve from level $n$ to level $n'$ both $A_n^q$ and $A_{n'}^q$
must be large, but if $A_n^q=\delta_{n,q}$, then $A_{n'}^q=0$
for all $n'\ne n$. Hence boundary of the probability
distribution spreading should be determined from the
condition of localization
of the QE eigenfunctions (the condition
$n \hbar\delta\omega\ge V_0$ was satisfied for the
boundary of the probability distribution spreading in Fig. 6).
The steep decay of the probability distribution in Fig. 6
in the quasiclassical approximation corresponds to absence of 
tunneling
between invariant curves. A qualitative comparison of permeabilities
of the invariant curves in the case $\delta\omega\ne 0$ with
permeabilities of the "quantum separatrices" in the case
$\delta\omega = 0$ (see Fig. 2) indicates that the latter are much
more transparent (compare also with Ref. [1]).

In conclusion we would like to do the following remark.
The nonperturbed Hamiltonian $\hat H_0$ yields an unlimited (from 
above)
set of Landau levels with equal distance $\hbar\omega_c$ between 
them.
When we add a monochromatic wave with the same frequency
$\omega=\omega_c$, a particle can pass to other levels and
an interesting question arises: is the diffusion limited or not?
The answer depends on how the perturbation depends on coordinate $x$.
If the dependence is linear then quantum states are
delocalized and the particle goes to infinity in the limit
$t\rightarrow\infty$ (as in the classical case) [10, 11].
As follows from the results of our work,
thanks to nonlinearity of the potential $V_0\cos(kx-\omega t)$ the 
quantum
states (as classical ones) turn out to be localized. However,
the peculiarity of localization in the latter case is that the sharp 
decay
of the probability distribution occures not everywhere, but only
at the boundaries of the quantum resonance cells.

In the next paper we will focus on studying destruction of the
quantum resonances under the influence of nonresonant terms in the
Hamiltonian (1) when the amplitude of perturbation is not small.

This research was made possible thanks to financial support
from the High School Commitee of Russia (Grant No. 95-0-5.5-63),
Russian Foundation for Basic Research (Grant No. 95-02-05620
and Grant No. 96-02-18067a) and Grant from INCAS (Grant No. 97-2-15).

\begin{figure}
\caption{Contour plots in the coherent state representation for
the eigenstates with eigenvalues at the top and bottom of the QE 
spectrum
$N=100$, $h=0.52$ (a). Poincare surfaces of section for one cell
for the same parameters of the Hamiltonian (b). }
\label{Fig. 1}
\end{figure}
\begin{figure}
\caption{Probability distribution for seven quantum cells,
$h=0.6$, $V_0=0.1$, the number of levels $N=432$,
(a) $t=4*10^5$, (b) $t=7*10^5$.
The initial state was placed in the center of the first cell.}
\label{Fig. 2}
\end{figure}
\begin{figure}
\caption{The average probability distribution
for (a) the first cell $P_1(t)$ and (b) the second cell $P_2(t)$.
The parameters and the initial conditions are the same as in Fig. 2.}
\label{Fig. 3}
\end{figure}
\begin{figure}
\caption{Time-averaged probability distribution for two values of
the effective Plank's constant $h$
versus cell number. The parameters are the same as in Figs. 2, 3.}
\label{Fig. 4}
\end{figure}
\begin{figure}
\caption{Oscillations of the penetration coefficient
from second to the subsequent cells, $t=4*10^4$, $N=100$
(three cells), $V_0=0.1$.}
\label{Fig. 5}
\end{figure}
\begin{figure}
\caption{Probability distribution for the near resonance case
$\delta\omega=0.003$, $t=4*10^6$, $h=0.52$, $V_0=0.1$,
the number of levels $N=100$. The initial state in the form
$C_n(0)=\delta_{n,n_0}$ was situated at the level $n_0=6$.}
\label{Fig. 6}
\end{figure}
\end{document}